# Evidence for the existence of *Kondo coupled resonant modes* in heavy fermions

L. M. Holanda [1], J. M. Vargas [1], C. Rettori [1], S. Nakatsuji [2], K. Kuga [2], Z. Fisk [3], S.B. Oseroff [4] and P. G. Pagliuso [1]
[1] *Instituto de Física "Gleb Wataghin", UNICAMP, Campinas-SP, 13083-970, Brazil.*
[2] *Institute for solid State Physics (ISSP), University of Tokyo, Kashiwa 277-8581, Japan.*
[3] *University of California, Irvine, California 92697-4573, U.S.A.*
[4] *San Diego State University, San Diego, California 92182, U.S.A.*

**Electron Spin Resonance (ESR) can microscopically probe both conduction electrons (*ce*) and local moment (LM) spin systems in different materials. A *ce* spin resonance (CESR) is observed in metallic systems based on light elements or with enhanced Pauli susceptibility. LM ESR is frequently seen in compounds with paramagnetic ions and localized $d$ or $f$ electrons. Here we report a remarkable and unprecedented ESR signal in the heavy fermion (HF) superconductor $\beta$-$YbAlB_4$[1] which behaves as a CESR at high temperatures and acquires characteristics of the $Yb^{3+}$ LM ESR at low temperature. This dual behavior in same ESR spectra strikes as an *in situ* unique observation of the Kondo quasiparticles giving rise to a new ESR response called Kondo coupled resonant mode (KCRM). The proximity to a quantum critical point (QCP) may favor the observation of a KCRM and its dual character in $\beta$-$YbAlB_4$ may unveil the $4f$-electrons nature at the QCP.**

One of the most important and heavily studied problem in condensed matter physics involves the microscopic understanding of how localized $f$-electrons at high$-T$ evolve to itinerant heavy quasi-particles in a low$-T$ metallic state. The fundamental mechanism of this evolution lies at the heart of heavy-electron physics and depends on the Kondo coupling between the localized $f$-electrons and the conduction electrons (*ce*).[2]

In principle, ESR would be one of the main techniques to bring insights to this problem since it could probe directly the $f-$electrons of Kondo ions and their interaction with the *ce*. However, for many years, it was generally accepted that the ESR lines of a Kondo ion such as $Yb^{3+}$ or $Ce^{3+}$ would broaden dramatically at low-$T$ avoiding their observation. Only few years ago, a breaking-through ESR signal was found in a Kondo lattice compound, the antiferromagnetic(AFM) heavy-fermion (HF) $YbRh_2Si_2$[3]. After this first observation, few others HF systems were found to present ESR lines[4, 5] and theoretical models[6, 7, 8] were proposed to explain the origin of such unexpected signals. However, the remarkable ESR signal reported here for $\beta$-$YbAlB_4$ comes to challenge any previous understanding about the origin of these signals that was believed to exist so far.

$\beta$-$YbAlB_4$ is the first reported Yb-based heavy-fermion superconductor (HFS)($T_N$ = 70 mK).[1] It is a new morphology of the previously known $\alpha$-$YbAlB_4$ phase which is a paramagnetic HF at low-$T$.[9] In contrast, for the $\beta$-$YbAlB_4$ phase, the low-$T$ superconducting state emerges from a non-fermi-liquid (NFL) normal state associated to quantum criticality.[1] Perhaps not coincidentally, $YbRh_2Si_2$, the first HF to show an ESR signal[3, 10] also presents pronounced NFL behavior when its AFM state ($T_N$ = 70 mK) is tuned towards a quantum critical point (QCP) by magnetic field.[11, 12]

Figure 1a presents the room-$T$ X-Band ESR spectra for fine powder of $\alpha$- and $\beta$-$YbAlB_4$ compounds and Fig. 1b the high-$T$ and low-$T$ X-Band ESR spectra for fine powder of $\beta$-$YbAlB_4$. The best fits to the data for $\alpha$-$YbAlB_4$ yield $g$ = 2.1(1) and linewidth $\Delta H$ = 700(70) Oe at high-$T$, and for $\beta$-$YbAlB_4$ $g$ = 2.34(8) and $\Delta H$ = 260(20) Oe at high-$T$ and a $g$ = 2.98(6) and $\Delta H$ = 150(10) Oe at low-$T$.

In the low-$T$ spectra of $\beta$-$YbAlB_4$ the arrows clearly indicate the presence of the hyperfine lines associated with the $^{171}Yb$ (I = 1/2) isotope. Using the field position of the $^{171}Yb$ hyperfine lines, extracted from the low-$T$ spectra of Fig. 1b and the Breit-Rabi formula[13], we obtain the hyperfine constant $^{171}A \approx 1300$ Oe which is of the order of typical values found for $^{171}Yb$ in low symmetry systems.[14, 15] The observation of these hyperfine lines is an irrefutable indication that the ESR spectra found for $\beta$-$YbAlB_4$ acquires the characteristic of the $Yb^{3+}$ ions at low-$T$. We should mention that no hyperfine lines were observed at higher-$T$.

The $T$-dependence of the ESR parameters, $g$-value, $\Delta H$ and intensity, for the fine powder of both $\alpha$- and $\beta$-$YbAlB_4$ compounds is shown in Figure 2. For $\beta$-$YbAlB_4$ and $T \lesssim 100$ K Fig 2a shows that the $g$-value increases as $T$-decreases, i.e., the resonance shift toward lower fields (see Fig. 1b) and at $T$ = 4.2 K the ESR line reaches $g \sim 3.0$. This $g$-value is close to the $g$-value found for $Yb^{3+}$ Kramers doublets in different crystal symmetries.[3, 4, 10, 14, 16] In contrast, the ESR line of $\alpha$-$YbAlB_4$ presents a nearly $T$-independent $g \sim 2.3$ down to $\approx 40$ K.

A striking difference between the $T$-dependence of $\Delta H$ is also verified for the $\alpha$- and $\beta$-$YbAlB_4$ compounds (see Fig. 2b). For $\beta$-$YbAlB_4$ $\Delta H(T)$ shows a weak non-monotonic increase as a function of temperature which results in an average linewidth broadening of $\approx$ 0.4 Oe/K in the whole temperature range. This rate is much smaller than the linear Korringa-rate found for the ESR line observed at low-$T$ for $YbRh_2Si_2$[3, 10] In contrast, for $\alpha$-$YbAlB_4$ $\Delta H(T)$ increases dramatically with

decreasing-$T$ which difficult the observation of the resonance for $T \lesssim 40$ K.

Lastly Figure 2c displays a very important and conclusive result. The ESR intensity obtained from the double integral of the ESR spectra is nearly $T$-independent in the whole studied $T$-range for both samples. This is a typical CESR behavior since the *ce* present a $T$-independent Pauli magnetic susceptibility and it is in dramatic contrast to what is expected for the $T$-dependence of a LM ESR intensity as in $YbRh_2Si_2$[3, 10] (Curie-like behavior, see Fig 2c). Thus, these results show that the ESR signal found in the NFL normal state of the HFS $\beta$-$YbAlB_4$ displays the typical behavior of CESR which acquires at low-$T$ characteristics of $Yb^{3+}$ LM ($g$-value and hyperfine splitting). On the other hand, the paramagnetic FL $\alpha$-$YbAlB_4$ presents an ESR signal that behaves as CESR in the whole studied $T$-range ($T$-independent $g$-value and ESR intensity) apart from the dramatic line broadening at low-$T$ and the $g$-value $\sim 2.3$ reasonably larger than $g = 2$ for free electrons.

The above description concerning the behavior of the ESR spectra in the two phases is further confirmed when the anisotropy of the ESR spectra for both phases is investigated.

Figure 3 display the angular dependence of the $g$-values for crystals of both $YbAlB_4$ phases at different temperatures. The $g$-value is isotropic and $T$-independent for $\alpha$-$YbAlB_4$, as expected for a CESR. However, for $\beta$-$YbAlB_4$, the $g$-value is isotropic at room-$T$ but becomes clearly anisotropic at $T = 4.2$ K, as it would be expected for an ESR signal arising from a $Yb^{3+}$ LM Kramers doublet in orthorhombic symmetry.

The striking and unique dual behavior observed in the same ESR spectra of $\beta$-$YbAlB_4$ ($T_c = 80$ mK), which behaves as a CESR at high-$T$ and acquires characteristics of $Yb^{3+}$ LM at low-$T$, associated to the ESR results found for $\alpha$-$YbAlB_4$, $YbRh_2Si_2$ and other HF compounds,[4, 5] allow us to propose a qualitative scenario that may explain the origin of the ESR signal in HF systems and, more fundamentally, contribute to a further understanding of the actual character of the $4f$ electrons at a QCP.

In order to build up such a scenario one has to go back to the classical transmission ESR (TESR) experiments in Ag:Dy and Ag:Er alloys that, respectively, allowed the simultaneous observation at low-$T$ of the CESR and the $Er^{3+}$ and $Dy^{3+}$ ESR $4f$ LM in their Kramers doublet ground state.[20] In these experiments it was shown that, as $T$-decreases, the CESR shifts to lower field showing an increase in the $g$-value which was found to be proportional to the *dc*-magnetic susceptibility of the LM. The LM ESR showed a $T$-independent $g$-shift proportional to the *ce* Pauli susceptibility (Knight-shift). However, at $T = 1.5$ K it was possible to observe that the LM ESR (g $\approx 7.6$ for $Dy^{3+}$) and the CESR line (g $\approx 2.4$) were well separated ESR signals. Furthermore, as a function of $T$ the two signals evolved accordingly to their individual characteristics. For instance, the intensity of the LM Kramers doublet ESR decreases dramatically with increasing-$T$ while the CESR could be followed to much higher-$T$.[20]

In the case of Kondo ions such as $Yb^{3+}$ it is known that the exchange coupling between the $4f$ LM and the *ce*, $J_{fs}$, is much stronger than that for non-Kondo earth-rare ions such as $Dy^{3+}$ and, in some compounds, these two spin system ($4f$ and *ce*) may be strongly hybridized.[2] Thus, it is entirely possible that for HF systems, the two independent ESR responses mentioned above become a unique $4f$-*ce* strongly coupled ESR mode that we named *Kondo coupled resonant mode*(KCRM). The KCRM is a new ESR response that possess dual nature, a CESR and/or LM ESR, depending on the strength of $J_{fs}$. For instance, HF systems with large Kondo energy scale ($T_K$) situated in the FL region of a Doniach-like phase diagram[2] would tend to present a CESR-like KCRM that may be observable depending on the material properties, e.g., metals with low *ce* spin-flip scattering (light metals with small *spin-orbit* coupling) and/or metals with enhanced Pauli magnetic susceptibility.[21] Typical FL HF are, for instance, $YbInCu_4$,[22] and $YbAgCu_4$,[23]. In these systems the KCRM would be expected to be CESR-like and should not be observed due to the large *ce* spin-flip scattering expected for the In, Cu and Ag elements. However, $\alpha$-$YbAlB_4$ presents a CESR-like KCRM because the *ce* spin-flip scattering is normally expected to be small for the light B and Al elements.

On the other hand, HF systems with small $T_K$ which may show magnetic ordering at low-$T$ (eg. $YbRh_2Si_2$, $YbIr_2Si_2$ and CeRuPO) would present in their paramagnetic state a LM-like KCRM that may be observable depending on the $f$-electrons spin-lattice relaxation rate involving the *ce* (Korringa-rate, bottleneck/dynamic effects), crystal field excited states, phonons and magnetic correlations. In particular, the strong bottleneck regime may favor the observation of LM-like KCMR.[3, 4, 5, 10, 24]

Moreover, HF systems near a QCP may present a KCRM that share the nature of both, the LM-like and CESR-like ESR signal. We argue that this is the case for the amazing ESR signal observed in $\beta$-$YbAlB_4$ where, as a function of $T$, the ESR signal presents both behaviors. Furthermore, the fact that this ESR signal presumably arises from Kondo $4f$-*ce* coupled quasi-particles that captures the $Yb^{3+}$ ionic characteristic at low-$T$, indicates that the scenario of local quantum criticality[25] is the one that more properly describe the behavior of the $4f$ electrons near a QCP.[2, 25, 26]

In summary, we report a new remarkable ESR signal in the HFS $\beta$-$YbAlB_4$ phase ($T_c = 80$ mK). This ESR signal has the unique behavior of CESR at high-$T$ and acquires characteristics of a LM $Yb^{3+}$ ESR at low-$T$. This dual nature was never observed before and is not found in the polymorph $\alpha$-$YbAlB_4$ phase which presents a CESR-like signal only. We argue that the $Yb^{3+}$ ionic character acquired by the ESR signal at low-$T$ in $\beta$-$YbAlB_4$ indicates that the $Yb^{3+}$ $4f$ electrons show a strong localized character at the QCP.

## I. METHODS

Single crystals of $\alpha$- and $\beta$-$YbAlB_4$ phases were grown from Al-flux as described previously.[1, 9] The crystals structures and phase purity were checked by x-ray powder diffraction. For the $\beta$-phase, the typical crystals size were ~0.5 x 0.5 x 0.05 $mm^3$ and their typical mass were less than ~ 0.1 mg. In contrast, the $\alpha$-phase crystals were much larger with typical dimensions of ~1.0 x 1.0 x 0.5 $mm^3$ and masses of ~ 4.0 mg. Most of the ESR data for both phases were taken using powdered crystals in order to increase the signal-to-noise ratio. To evaluate the anisotropy of the $\beta$-$YbAlB_4$ ESR spectra, ~ 40 oriented platelet-likes single-crystals were glued on several flat plastic surfaces and mounted in a form of sandwiches with the crystals $c$-axis perpendicular to the surface. The total mass of the used crystals was about ~ 4.0 mg. For the $\alpha$-phase, a 4.0 mg single crystal was used for the anisotropy studies. The ESR spectra were taken in a Bruker X-band (9.5 GHz) spectrometer using appropriated resonators and $T$-controller systems. Dysonian ESR lineshapes were observed for both samples in the whole $T$-range which corresponds to a microwave skin depth smaller than the size of the crystals.[27] The ESR signal for both samples was calibrated at room temperature using the ESR signal of a strong pith standard with 4.55 X $10^{15}$ spins/cm. The number of resonating spins extracted from this calibration for both samples were found to be of the same order of the number of Yb atoms within the samples skin depth.

## II. ACKNOWLEDGMENTS

This work is supported by FAPESP, CNPq and CAPES (Brazil), by NSF (USA), by Grant-in-Aid for Scientific Research (No. 21684019) from the JSPS and by Grant-in-Aid for Scientific Research on Priority Areas (No. 17071003) from MEXT, Japan.

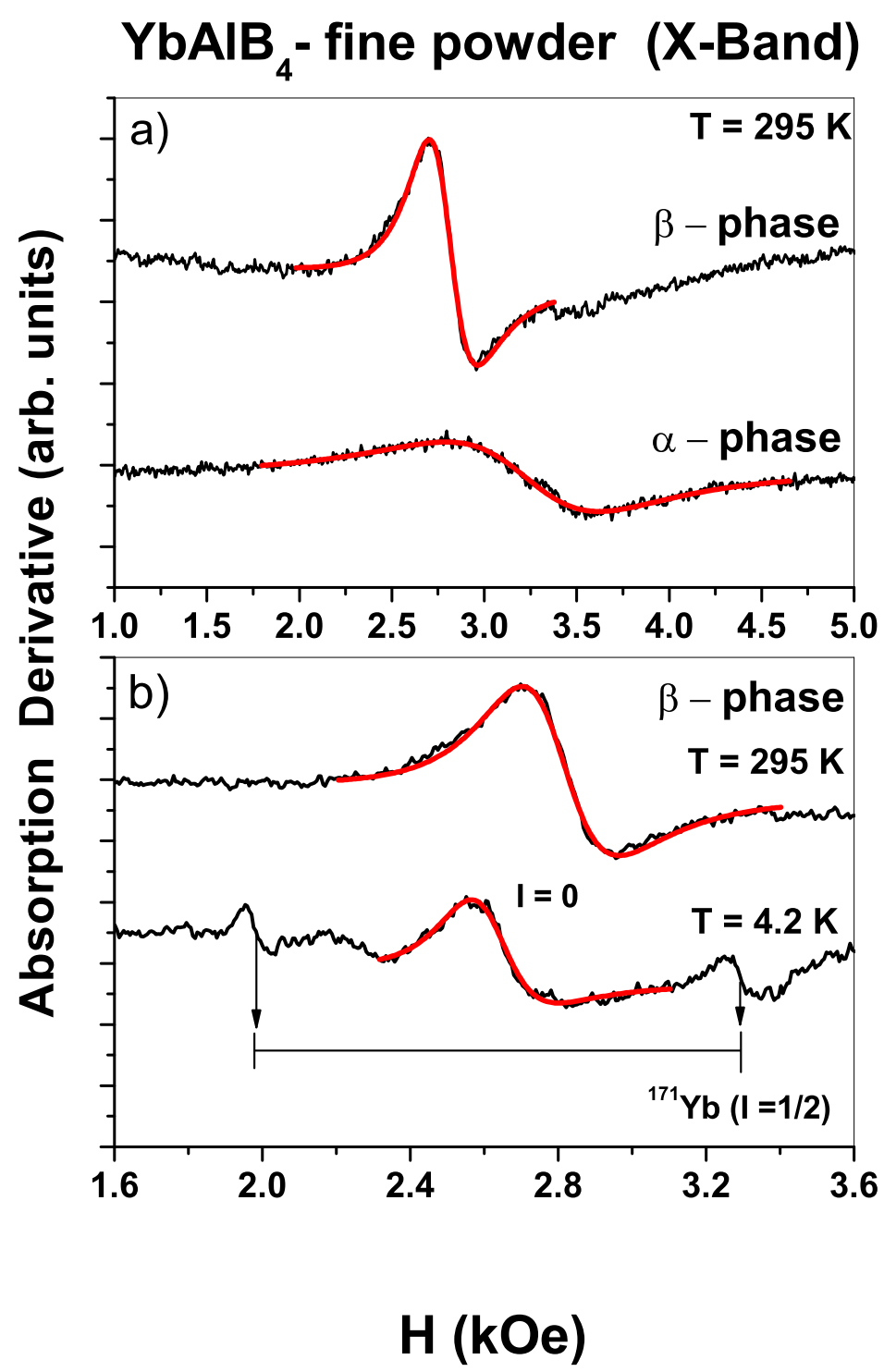

FIG. 1: (color online) a) Room-$T$ X-Band ESR spectra for a fine powder of $\alpha$- and $\beta$-$YbAlB_4$ compounds and b) high-$T$ and low-$T$ X-Band ESR spectra of the fine powder of $\beta$-$YbAlB_4$.The solid lines represents the best fits to the spectra using a Dysonian lineshape. Natural Yb has $\sim$ 70 % of $^{170}$Yb (I = 0), $\sim$ 14 % of $^{171}$Yb (I = 1/2) and $\sim$ 16 % of $^{173}$Yb (I = 5/2) isotopes. The hyperfine lines of $^{173}$Yb are usually more than 2 times less intense than the lines associated with $^{171}$Yb (I = 1/2) and they are not obviously observable in the spectra of Fig. 1b.

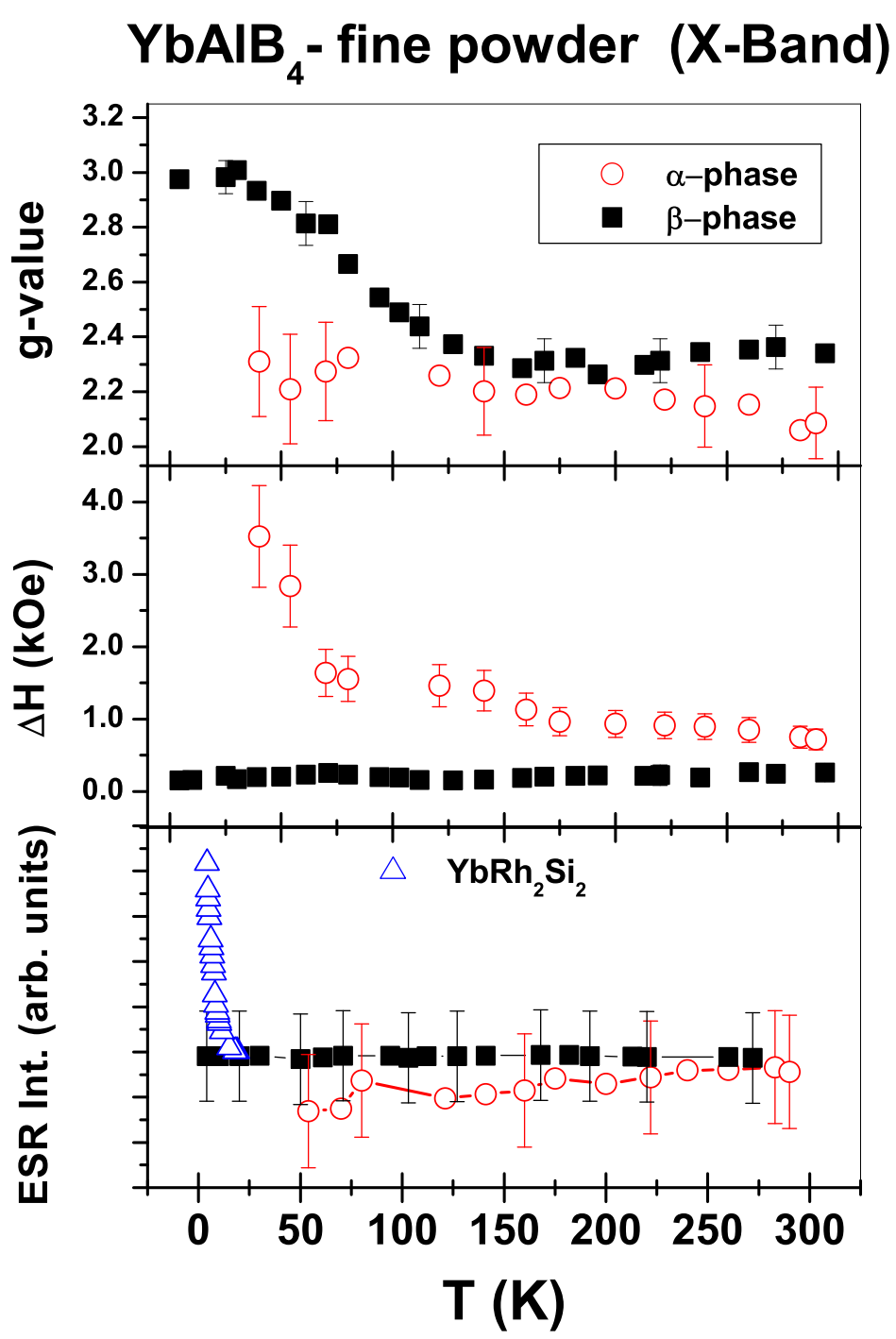

FIG. 2: (color online) The temperature dependence of the ESR parameters, $g$-value, $\Delta H$ and intensity for fine powder of both $\alpha$- and $\beta$-$YbAlB_4$ compounds. The broadening of the ESR line of $\alpha$-$YbAlB_4$ at low-$T$ is typical of ESR lines in the presence of spin-spin interaction and it may represent the $ce$ electron-electron interaction in the FL regime of phase $\alpha$-$YbAlB_4$ captured from the perspective of a CESR-like mode.

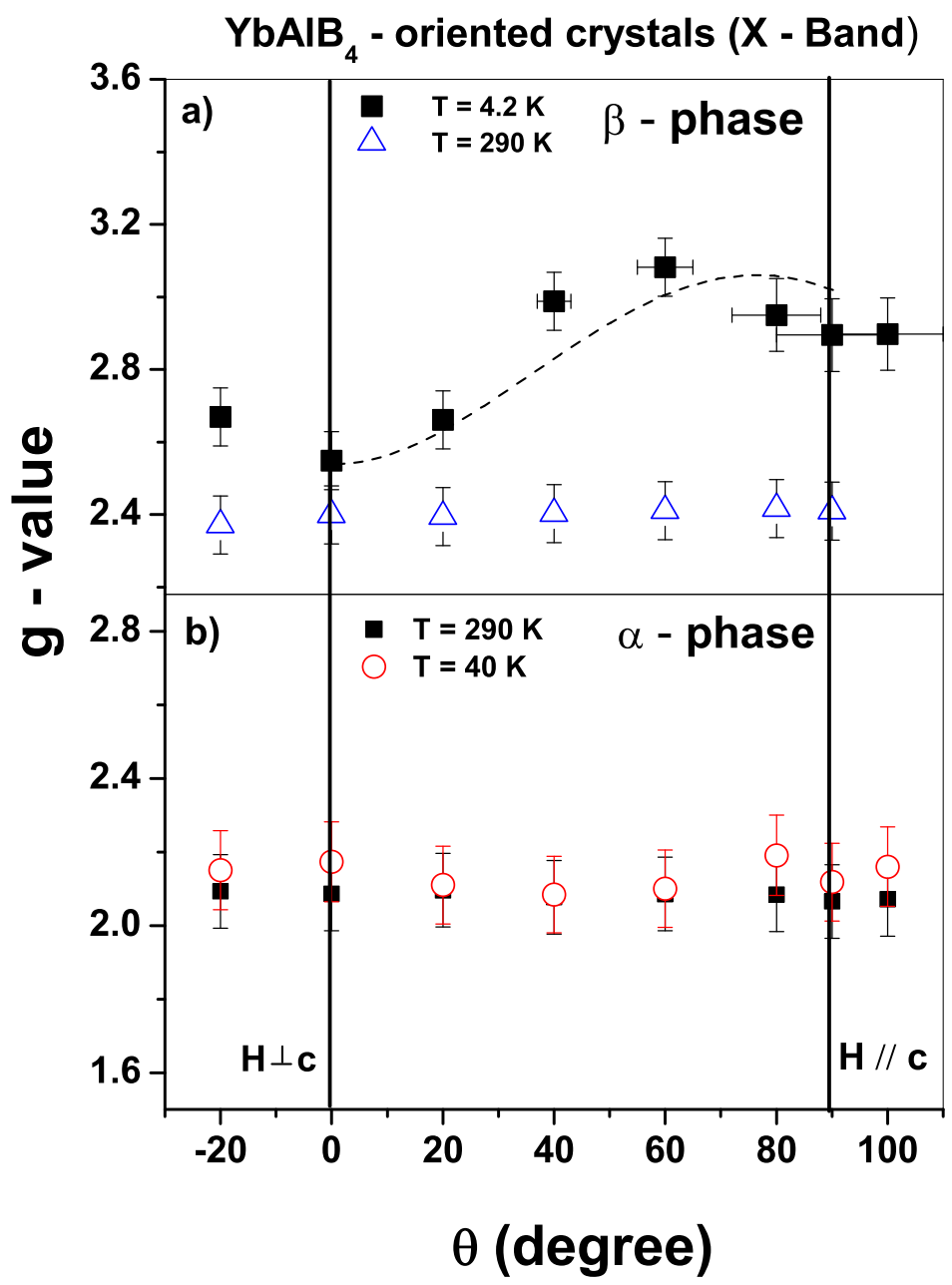

FIG. 3: (color online) $g$-values as a function of angle for oriented crystals of both phases at different temperatures($T$ = 290 K and $T$ = 4.2 K for $\beta$-YbAlB$_4$ and $T$ = 290 K and $T$ = 40 K for $\alpha$-YbAlB$_4$. The dashed line is just a guide to the eye. The observed anisotropy at $T$ = 4.2 K for $\beta$-YbAlB$_4$ shows the largest $g$-value when $H$ is applied along the $c$-axis, consistent with the largest magnetic susceptibility measured for this field orientation.[1, 9] Interestingly, the $g$-value anisotropy of the $\beta$-YbAlB$_4$ phase is in contrast to that found for YbRh$_2$Si$_2$,[3, 10] where the largest $g$-value is found for $H$ perpendicular to the $c$-axis. This change in the single ion anisotropy is probably associate with a change in the symmetry of the crystal field ground state which may be also relevant for driving the compound ground state from AFM (YbRh$_2$Si$_2$) to HFS $\beta$-YbAlB$_4$.[17, 18, 19]Nevertheless, for both $\beta$-YbAlB$_4$ and YbRh$_2$Si$_2$ it is evident that the observed ESR signal is reflecting the Yb$^{3+}$ single ion LM anisotropy in that particular crystal symmetry.